\documentclass[aps,prb]{revtex4}
\usepackage{amsmath}
\usepackage{graphicx}
\usepackage{amsmath}

\begin{document}
\pagestyle{empty}
\title{  Casimir-Lifshitz forces and radiative heat
transfer between moving bodies.}

\author{A.I. Volokitin$^{1,2}$\footnote{Corresponding author.
\textit{E-mail address}:alevolokitin@yandex.ru}    and B.N.J.
Persson$^2$}  \affiliation{$^1$ Samara State Technical University,
443100 Samara, Russia} \affiliation{$^2$Institut f\"ur
Festk\"orperforschung, Forschungszentrum J\"ulich, D-52425,
Germany}
\begin{abstract}
Recently Philbin \textit{et al} \{New J. Phys.  11 (2009) 033035;
arXiv:0904.2148v3 [quant-ph], 2009\} have presented a new theory of the
van der Waals friction. Contrary to previous theories they
claimed that there is no ``quantum friction'' at zero temperature.
We show that this theory is incorrect.

\end{abstract}
\maketitle

\section{Introduction}

All bodies are surrounded by a fluctuating electromagnetic field
due to the thermal and quantum fluctuations of the charge  and
current density inside the bodies. Outside the bodies this
fluctuating electromagnetic field exists partly in the form of
propagating electromagnetic waves and partly in the form of
evanescent  waves. The theory of the fluctuating electromagnetic
field was developed by Rytov \cite{Rytov53,Rytov67,Rytov89}. A
great variety of phenomena such as Casimir-Lifshitz forces
\cite{Lifshitz54},  near-field radiative heat transfer
\cite{Polder}, and friction forces
\cite{Volokitin99,VolokitinRMP07,VolokitinUFN07,Volokitin08} can
be described using this theory.

Lifshitz \cite{Lifshitz54} used the Rytov's theory to formulate a
very general theory of the dispersion interaction using
statistical physics and macroscopic electrodynamics. The
Lifshitz theory  provides a common tool to deal with dispersive
forces in different fields of science (physics, biology,
chemistry) and technology.

The Lifshitz theory is valid for systems at thermal
equilibrium. At present there is an interest in the study of
systems out of the thermal equilibrium (see \cite{Pitaevskii08}
and reference therein), in particular in the connection with the
possibility of tuning the strength and sign of the interaction\cite{Antezza05,Antezza06}.
Such systems also present a way to
explore the role of thermal fluctuations, which usually are masked at thermal
equilibrium by the $T=0$ K component, which dominates the
interaction up to very large distances, where the interaction
force is very small. In Ref. \cite{Antezza05} the Casimir-Lifshitz force was
measured at very large distances, and it was shown that the thermal
effects on the Casimir-Lifshitz interaction agree with
the theoretical prediction. This measurement was
done out of thermal equilibrium, where thermal effects are
stronger.

Non-equilibrium thermal effects was also studied by Polder and Van
Hove\cite{Polder}, who calculated the heat-flux between two
parallel plates. At present there is an increasing interest in
near-field radiative heat
transfer\cite{Pendry99,Volokitin01a,Volokitin04,Volokitin03,Mulet01,Mulet02},
in the connection with the development of near-field scanning
thermal microscopy\cite{Kittel}. The existing studies are limited
mostly to the case when the interacting bodies are at rest. For
recent reviews of near-field radiative heat transfer between
bodies, which are at rest, see Refs.
\cite{Joulain05,VolokitinRMP07,VolokitinUFN07}.

Non-equilibrium effects always prevail for bodies moving relative
to each other. In Ref.\cite{Volokitin99} we used a dynamical
modification of the Lifshitz theory to calculate the friction
force between two parallel surfaces in relative motion (velocity
$V$). The calculation of the van der Waals friction is more
complicated than of the Casimir-Lifshitz force (and of the
radiative heat transfer), because it requires the determination of
the electromagnetic field between moving boundaries. The solution
can be found by writing the boundary conditions on the surface of
each body in the rest reference frame of this body. The relation
between the electromagnetic fields in the different reference
frames is determined by the Lorenz transformation. In
Ref.\cite{Volokitin99} the electromagnetic field in the vacuum gap
between the bodies was calculated to linear order in  $V/c$. These
linear terms corresponds to mixing of electromagnetic waves with
different polarizations. The waves with different polarization are
statistically independent. Thus after averaging of the stress
tensor over the fluctuating electromagnetic field, the mixing
terms will give a contribution to the friction force of order
$(V/c)^2$.  In Ref.\cite{Volokitin99} the mixing terms were
neglected, and the resulting formula for friction force is
accurate to order $(V/c)^2$.  The same approximation was used in
Ref.\cite{Volokitin01b} to calculate the frictional drag between
quantum wells, and in Refs.\cite{Volokitin03a,Volokitin03b} to
calculate the friction force between plane parallel surfaces in
normal relative motion.  In Ref. \cite{Volokitin06} the
correctness of the approach based on the dynamical modification of
the Lifshitz theory was confirmed (at least to linear order in the
sliding velocity $V$) by rigorous quantum mechanical calculations
(using the Kubo formula for friction coefficient). For recent
reviews of the van der Waals friction see Refs.
\cite{VolokitinRMP07,VolokitinUFN07}.

In Ref. \cite{Volokitin08}  we presented a of unified
approach to the Casimir-Lifshitz  forces and the radiative
heat transfer at nonequilibrium conditions, when the
bodies are at different temperatures, and move relative to
each other with an arbitrary velocity $V$.
In comparison with previous
calculations\cite{Volokitin99,Volokitin01b,Volokitin03a,Volokitin03b},
we did not make any approximation in the Lorentz
transformation of the electromagnetic field. Thus, we
could determine the field in one inertial reference frame, knowing
the same field in another reference frame, and our solution of
the electromagnetic problem was exact. Knowing the electromagnetic
field we calculated the stress tensor and the Poynting vector
which determined the Casimir-Lifshitz forces and the heat
transfer, respectively.  Upon going to the limit when one of the
bodies is rarefied we  obtained the interaction force and the heat
transfer for a small particle-surface configuration.

In the recent papers Philbin and Leonhardt
\cite{Philbin1,Philbin2} (henceforth refereed to as PL) calculated
the Casimir-Lifshitz forces due to electromagnetic fluctuations
between two perfectly flat parallel dielectric surfaces separated
by vacuum and moving parallel to each other. In
Ref.\cite{Philbin1} LP used Lifshitz theory
\cite{Lifshitz54,Lifshitz1961,Lifshitz1980} and considered only
the case of zero temperature. Lifshitz theory
\cite{Lifshitz54,Lifshitz1961,Lifshitz1980} also included the
effect of thermal radiation in his analysis. The
\textit{Casimir-Lifshitz effect} is therefore also taken to
describe forces that have a contribution from thermal radiation as
well as from the quantum vacuum. The formalism developed by
Lifshitz, however, cannot be used for plates at different
temperatures.  The general case of finite and different
temperatures was considered in Ref. \cite{Philbin2}.  In Ref.
\cite{Philbin2} LP used the same approach as in Ref.
\cite{Volokitin08}, which is based on a dynamical modification of
the Rytov's  theory. Theory from Ref. \cite{Philbin2} contains as
a limiting case the theory from Ref. \cite{Philbin1}. For the
contributions to the Casimir-Lifshitz forces resulting from
thermal fluctuations LP obtained the same results as in Ref.
\cite{Volokitin08}. However at zero temperature LP obtained a
result which contradicts a substantial body of earlier results
\cite{VolokitinRMP07,VolokitinUFN07,Volokitin99,Volokitin08,Pendry97}.
Their conclusion was that, at zero temperature, where only quantum
fluctuations occur, friction is precisely zero. In this paper we
argue for the correctness of the earlier results and point to the
errors in the reasoning of PL.

\section{Basic results}

\begin{figure}
\includegraphics[width=0.45\textwidth]{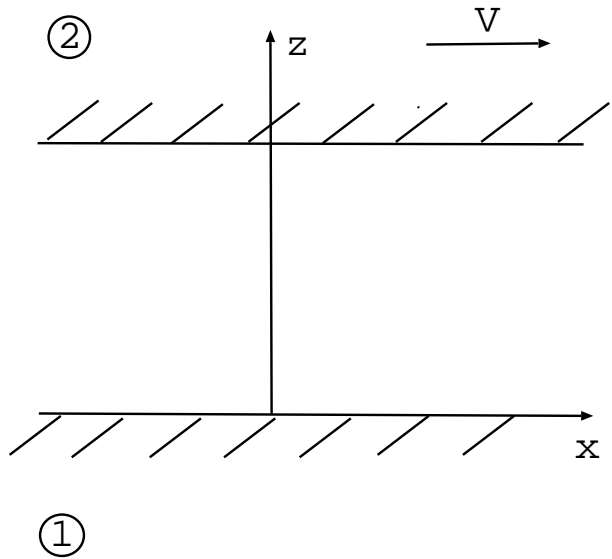}
\caption{\label{Fig1} Two semi-infinite bodies with plane parallel
surfaces separated by a distance $d$. The upper solid moves
parallel to  the other solid with the velocity $V$. }
\end{figure}

We consider two semi-infinite solids having flat parallel surfaces
separated by a distance $d$ and moving with velocity $V$ relative
to each other, see Fig. \ref{Fig1}.
We introduce two coordinate systems $K$ and $K^{\prime }$ with
coordinate axes $xyz$ and $x^{\prime }y^{\prime }z^{\prime }$. In
the $K$ system body \textbf{1} is at rest while body \textbf{2}
moves with the velocity $V$ along the $x-$ axis. The $xy$ and
$x^{\prime }y^{\prime }$ planes are in the surface of body
\textbf{1}, and the $x$ and $ x^{\prime}$- axes are parallel.
The $z$ and $z^{\prime}-$ axes point toward body \textbf{2}.
In the $K^{\prime}$ system body \textbf{2} is at rest while body
\textbf{1} is moving with velocity $-V$ along the $x-$ axis.

The force which acts  on the surface of body \textbf{1} can be
calculated from the Maxwell stress tensor $\sigma_{ij}$, evaluated
at $z=0$:
\[
\sigma _{ij} =\frac 1{4\pi }\int_0^\infty d\omega \int
\frac{d^2q}{(2\pi)^2} \Big[ <E_iE_j^*> + <E_i^*E_j> + <B_iB_j^*> +
<B_i^*B_j>.
\]
\begin{equation}
 - \delta_{ij}(<\mathbf{E\cdot E^*}> + <\mathbf{B\cdot B^*}>)\Big] _{z=0}   \label{3one}
\end{equation}

According to Ref. \cite{Volokitin08} the $x$-component of the
force is given by
\[
F_x = \sigma_{xz} =\frac \hbar {8\pi ^3}\int_0^\infty d\omega
\int_{q<\omega /c}d^2q\frac{q_x}{|\Delta|^2}[(q^2 - \beta kq_x)^2
+ \beta^2k_z^2q_y^2]
\]
\[
\times  [(q^2 - \beta kq_x)^2(1-\mid R_{1p}\mid ^2)(1-\mid
R_{2p}^{\prime }\mid ^2)|D_{ss}|^2
\]
\[
+\beta^2k_z^2q_y^2(1-\mid R_{1p}\mid ^2)(1-\mid R_{2s}^{\prime
}\mid ^2)|D_{sp}|^2 + (p\leftrightarrow s)] \left(
n_2(\omega^{\prime})-n_1(\omega )\right)
\]
\[
+\frac \hbar {2\pi ^3}\int_0^\infty d\omega \int_{q>\omega
/c}d^2q\frac{q_x}{|\Delta|^2}[(q^2 - \beta kq_x)^2 +
\beta^2k_z^2q_y^2] e^{-2\mid k_z\mid d}
\]
\[
\times [(q^2 - \beta kq_x)^2
\mathrm{Im}R_{1p}\mathrm{Im}R_{2p}^{\prime}|D_{ss}|^2-
\beta^2k_z^2q_y^2 \mathrm{Im}R_{1p}\mathrm{Im}
R_{2s}^{\prime}|D_{sp}|^2
\]
\begin{equation}
+ (p\leftrightarrow s)]\left( n_2(\omega^{\prime})-n_1(\omega
)\right),   \label{2three}
\end{equation}
where $R_{1p(s)}=R_{1p(s)}(\omega,q)$ and
$R_{2p(s)}^{\prime}=R_{2p(s)}(\omega^{\prime},q^{\prime})$ are the
reflection amplitudes for surfaces \textbf{1} and  \textbf{2} for
the $p(s)$ - polarized electromagnetic field,  respectively,
$\mathbf{q}=(q_x,q_y)$, $k_z=((\omega/c)^2-q^2)^{1/2}$,
$\mathbf{q}^{\prime}=(q_x^{\prime},q_y),\,q_x^{\prime}=(q_x -
\beta k)\gamma,\,\omega^{\prime}= (\omega - Vq_x)\gamma,\,
\gamma=1/\sqrt{1-\beta^2},\,\beta = V/c$, $k=\omega/c$,
\[
q^{\prime}=\gamma\sqrt{q^2-2\beta kq_x+\beta^2(k^2-q_y^2)},
\]
\[
D_{pp} = 1 - e^{2ik_zd}R_{1p}R_{2p}^{\prime},\, D_{ss} = 1 -
e^{2ik_zd}R_{1s}R_{2s}^{\prime},
\]
\[
D_{sp} = 1 + e^{2ik_zd}R_{1s}R_{2p}^{\prime},\, D_{ps} = 1 +
e^{2ik_zd}R_{1p}R_{2s}^{\prime},
\]
\[
\Delta = (q^2 - \beta kq_x)^2D_{ss}D_{pp} +
\beta^2k_z^2q_y^2D_{ps}D_{sp},
\]
\[
n_i(\omega )=\frac 1{e^{\hbar \omega /k_BT_i}-1},
\]
where $T_1$ and $T_2$ are the temperatures for bodies \textbf{1}
and \textbf{2}, respectively. The symbol $(p\leftrightarrow s)$
denotes the terms which can be obtained from the preceding terms
by permutation of the indexes $p$ and $s$. The first term in Eq.
(\ref{2three}) represents the contribution to the friction from
propagating waves ($q<\omega /c$), and the second term  from the
evanescent waves ($q>\omega /c$). If in Eq. (\ref{2three})  one
neglects the terms of the order $\beta^2$ then the contributions
from waves with $p$- and $s$- polarization will be separated. In
this case Eq. (\ref{2three}) is reduced to the formula obtained in
Ref.\cite{Volokitin99}. Thus, to the order $\beta^2$ the mixing of
waves with different polarization can be neglected, what agrees
with the results obtained in Ref.\cite{Volokitin99}. At $T=0$ K
the propagating waves do not contribute to friction but the
contribution from evanescent waves is not equal to zero. Taking
into account that $n(-\omega)= -1 - n(\omega)$ from Eq.
(\ref{2three}) we get the friction mediated by the evanescent
electromagnetic waves at zero temperature (in literature this type
of friction is denoted as quantum friction \cite{Pendry97})
\[
F_x = - \frac \hbar {\pi ^3}\int_0^\infty dq_y \int_0^\infty dq_x
\int_0^{q_xV} d\omega \frac{q_x}{|\Delta|^2}[(q^2 - \beta kq_x)^2
+ \beta^2k_z^2q_y^2] e^{-2\mid k_z\mid d}
\]
\begin{equation}
\times [(q^2 - \beta kq_x)^2
\mathrm{Im}R_{1p}\mathrm{Im}R_{2p}^{\prime}|D_{ss}|^2-
\beta^2k_z^2q_y^2 \mathrm{Im}R_{1p}\mathrm{Im}
R_{2s}^{\prime}|D_{sp}|^2 + (p\leftrightarrow s)].
\label{zerotemfr}
\end{equation}

In Ref. \cite{Volokitin08}  some additional terms were overlooked
in the equation for the
$z$-component of the force. The correct form of
this equation is given by \cite{Volokitin10}

\[
F_z = \sigma_{zz} =-\frac \hbar {4\pi ^3}\mathrm{Re}\int_0^\infty d\omega \int d^2q\frac{k_z}{\Delta}e^{2ik_zd}
\Big\{(q^2 -\beta kq_x)^2[R_{1p}R_{2p}^{\prime}D_{ss}
\]
\[
+ R_{1s}R_{2s}^{\prime}D_{pp}] -
\beta^2k_z^2q_y^2[R_{1p}R_{2s}^{\prime}D_{sp} +
R_{1s}R_{2p}^{\prime}D_{ps}]\Big\} [1 + n_1(\omega) +
n_2(\omega^{\prime})]
\]
\[
-\frac \hbar {16\pi ^3}\int_0^\infty d\omega \int_{q<\omega
/c}d^2q\frac{k_z}{|\Delta|^2}[(q^2 - \beta kq_x)^2 +
\beta^2k_z^2q_y^2]
\]
\[
\times  \{(q^2 - \beta kq_x)^2[(1-\mid R_{1p}\mid ^2)(1+\mid
R_{2p}^{\prime }\mid ^2)|D_{ss}|^2-(1\leftrightarrow 2,\omega
\leftrightarrow \omega^{\prime})]
\]
\[
+\beta^2k_z^2q_y^2[(1-\mid R_{1p}\mid ^2)(1+\mid R_{2s}^{\prime
}\mid ^2)|D_{sp}|^2 -(1\leftrightarrow 2,\omega \leftrightarrow
\omega^{\prime})]+ (p\leftrightarrow s)\} \left( n_1(\omega )-
n_2(\omega^{\prime})\right)
\]
\[
+\frac \hbar {4\pi ^3}\int_0^\infty d\omega \int_{q>\omega
/c}d^2q\frac{|k_z|}{|\Delta|^2}[(q^2 - \beta kq_x)^2 +
\beta^2k_z^2q_y^2] e^{-2\mid k_z\mid d}
\]
\[
\times \{(q^2 - \beta kq_x)^2[
\mathrm{Im}R_{1p}\mathrm{Re}R_{2p}^{\prime}|D_{ss}|^2-
(1\leftrightarrow 2,\omega \leftrightarrow \omega^{\prime})]-
 \beta^2k_z^2q_y^2[
\mathrm{Im}R_{1p}\mathrm{Re}
R_{2s}^{\prime}|D_{sp}|^2-(1\leftrightarrow 2,\omega
\leftrightarrow \omega^{\prime})]
\]
\begin{equation}
+ (p\leftrightarrow s)\}\left( n_1(\omega
)-n_2(\omega^{\prime})\right). \label{2five}
\end{equation}

 At $T_1=T_2=0$ K, Eq. (\ref{2five}) takes the form
\[
F_z =-\frac \hbar {4\pi ^3}\mathrm{Re}\left\{\int_0^\infty d\omega \int d^2q -
\int_{-\infty}^{\infty}dq_y\int_{0}^{\infty}dq_x
\int_0^{q_xV} d\omega\right\}\frac{k_z}{\Delta}e^{2ik_zd}
\]
\[
\Big\{(q^2 -\beta kq_x)^2[R_{1p}R_{2p}^{\prime}D_{ss}+ R_{1s}R_{2s}^{\prime}D_{pp}]
\]
\[
- \beta^2k_z^2q_y^2[R_{1p}R_{2s}^{\prime}D_{sp} +
 R_{1s}R_{2p}^{\prime}D_{ps}]\Big\}
 \]
  \[
 +\frac \hbar {4\pi ^3}\int_{-\infty}^{\infty}dq_y\int_{0}^{\infty}dq_x
\int_0^{q_xV} d\omega \frac{|k_z|}{|\Delta|^2}[(q^2 - \beta
kq_x)^2 + \beta^2k_z^2q_y^2] e^{-2\mid k_z\mid d}
\]
\[
\times \{(q^2 - \beta kq_x)^2[
\mathrm{Im}R_{1p}\mathrm{Re}R_{2p}^{\prime}|D_{ss}|^2-
(1\leftrightarrow 2,\omega \leftrightarrow \omega^{\prime})]-
 \beta^2k_z^2q_y^2[
\mathrm{Im}R_{1p}\mathrm{Re}
R_{2s}^{\prime}|D_{sp}|^2-(1\leftrightarrow 2,\omega
\leftrightarrow \omega^{\prime})]
\]
\begin{equation}
+ (p\leftrightarrow s)\}. \label{2six}
\end{equation}

The radiative energy transfer between the bodies is determined by
 the ensemble average of the Poynting's vector. In the case of  two plane parallel surfaces
$z$-component of the Poynting's vector is given by
\cite{VolokitinRMP07}
\[
\left\langle \mathbf{S}_{1z}(\mathbf{r})\right\rangle _\omega
=(c/8\pi )\left\langle \mathbf{E}(\mathbf{r})\times
\mathbf{B}^{*}(\mathbf{r} )\right\rangle _\omega +c.c.
\]
\begin{equation}
=\frac{ic^2}{8\pi \omega }\left\{ \langle \mathbf{E
}(\mathbf{r})\cdot \frac{d}{dz}\mathbf{E}^{*}(\mathbf{r})\rangle
-c.c\right\} _{z=0}. \label{poynting}
\end{equation}

According to Ref. \cite{Volokitin08} the heat flux across the
surface \textbf{1} is given by:

\[
S_1  =\frac {\hbar} {8\pi ^3}\int_0^\infty d\omega \int_{q<\omega
/c}d^2q\frac{\omega}{|\Delta|^2}[(q^2 - \beta kq_x)^2 +
\beta^2k_z^2q_y^2]
\]
\[
\times  [(q^2 - \beta kq_x)^2(1-\mid R_{1p}\mid ^2)(1-\mid
R_{2p}^{\prime }\mid ^2)|D_{ss}|^2
\]
\[
+\beta^2k_z^2q_y^2(1-\mid R_{1p}\mid ^2)(1-\mid R_{2s}^{\prime
}\mid ^2)|D_{sp}|^2 + (p\leftrightarrow s)] \left(
n_2(\omega^{\prime})-n_1(\omega )\right)
\]
\[
+\frac \hbar {2\pi ^3}\int_0^\infty d\omega \int_{q>\omega
/c}d^2q\frac{\omega}{|\Delta|^2}[(q^2 - \beta kq_x)^2 +
\beta^2k_z^2q_y^2] e^{-2\mid k_z\mid d}
\]
\[
\times [(q^2 - \beta kq_x)^2
\mathrm{Im}R_{1p}\mathrm{Im}R_{2p}^{\prime}|D_{ss}|^2-
\beta^2k_z^2q_y^2 \mathrm{Im}R_{1p}\mathrm{Im}
R_{2s}^{\prime}|D_{sp}|^2
\]
\begin{equation}
+ (p\leftrightarrow s)]\left( n_2(\omega^{\prime})-n_1(\omega
)\right).   \label{heat2}
\end{equation}

\section{Discussion and comparison with the results of PL}
In Ref. \cite{Philbin1}  PL used Lifshitz theory and considered
only the case of zero temperature. There are two variants  of
Lifshitz  theory. In the first variant the Maxwell stress tensor is
calculated using electromagnetic field which was calculated using
Rytov's theory. In the second variant the Maxwell stress tensor is
calculated using Green's functions of the electromagnetic field
which were calculated from Maxwell's equations. Both these
variants give the same results for forces. In Ref.  \cite{Philbin1}
PL used the second variant of the Lifshitz theory.  The general case
of finite and different temperatures was considered by PL in Ref.
\cite{Philbin2}.  The theory from Ref. \cite{Philbin2}  contains as
limiting case the theory from Ref. \cite{Philbin1}. In particular,
in both these theories the authors came to conclusion that there
is no lateral force on the plates in relative motion.
In Ref. \cite{Philbin2}  PL used the same approach as by Volokitin
\textit{et al} \cite{Volokitin08,Volokitin10} (henceforth referred
to as VP)  but they came to the opposite conclusion that there is
no ``quantum'' friction.  Between these two studies there is a
difference only in the technical details. VP calculated total
electromagnetic field in the rest reference frame  of surface
\textbf{1}. The total electromagnetic field contains the
contributions from quantum and thermal fluctuations in both
bodies. This electromagnetic field was used in the calculations of
the stress tensor and the Poynting's vector in the rest reference
frame  of surface \textbf{1}. LP divided the total stress tensor
into two contributions from surfaces \textbf{1} and \textbf{2}.
The contribution from surface \textbf{2} was first calculated in
the rest reference frame  of surface \textbf{2}. The contribution
from the surface \textbf{2} again was divided into two components
- one from quantum  and other from thermal fluctuations.
Separation of quantum-vacuum from thermal effect is achieved by
the identity
\begin{equation}
\mathrm{coth}\left (\frac{\hbar \omega}{2k_BT}\right )=
\mathrm{sgn}(\omega)+2\mathrm{sgn}(\omega)\left[\exp\left(\frac
{\hbar |\omega|}{k_BT_2}\right)-1\right]^{-1}\label{d0}
\end{equation}
where the first term gives the quantum-vacuum part and the the
second term, containing the Plank spectrum, gives the thermal
radiation part. The Lorentz transformation for the stress tensor was
used to obtain the contribution from thermal fluctuations in body
\textbf{2} in the rest reference frame  of surface \textbf{1}. The
integrand of this contribution contains factor
\begin{equation}
2\mathrm{sgn}(\omega^{\prime})\left[\exp\left(\frac {\hbar
|\omega^{\prime}|}{k_BT_2}\right)-1\right]^{-1}\label{d1a}
\end{equation}
As a result, for total contribution from thermal fluctuations in
both bodies, PL obtained exactly the same results as it was
obtained by VP. However, for the contribution from quantum
fluctuations PL proposed that the Lorentz transformation is not
valid, and arrived at the conclusion that the effect of zero-point
radiation for contribution from plate \textbf{2} can be obtained
by the following replacement of a factor in the integrand in the
expression for contribution from thermal fluctuations:
\begin{equation}
2\mathrm{sgn}(\omega^{\prime})\left[\exp\left(\frac {\hbar
|\omega^{\prime}|}{k_BT_2}\right)-1\right]^{-1}\rightarrow
\mathrm{sgn}(\omega) +
2\mathrm{sgn}(\omega^{\prime})\left[\exp\left(\frac {\hbar
|\omega^{\prime}|}{k_BT_2}\right)-1\right]^{-1} \label{d1}
\end{equation}
Contribution from plate \textbf{1} is given by a similar expression
as from plate \textbf{2}. As a result, for friction
force PL obtained expression which is similar to Eq.
\eqref{2three} but with replacement
\begin{equation}
(n_2(\omega^{\prime})-n_1(\omega))\rightarrow
(\mathrm{sgn}(\omega^{\prime})n_2(|\omega^{\prime}|)-n_1(\omega)).
\label{d2}
\end{equation}

For finite temperatures PL obtained the same contribution to
friction from thermal fluctuations as by VP. However, it is clear
from Eq. \eqref{d2} that at $T=0$ K the factor on the right side
of Eq. \eqref{d2} is equal to zero what leads PL to the conclusion
that there is no lateral force at zero temperature. PL claimed that
the vanishing of the lateral force at zero temperature can be
viewed as a consequence of the Lorentz invariance of the quantum
zero-point radiation--it has the same ``spectrum'' in every
inertial reference frame. However, instead of proving this PL just
postulated the existence of such an invariance. If the Lorentz
transformation is used also to calculate contribution to stress
tensor from quantum fluctuations in plate \textbf{2}, in the rest
reference frame of plate \textbf{1}, then instead of the factor
given by the right side of Eq. \eqref{d1}, in the integrand will
occur the factor
\begin{equation}
\mathrm{sgn}(\omega^{\prime}) +
2\mathrm{sgn}(\omega^{\prime})\left[\exp\left(\frac {\hbar
|\omega^{\prime}|}{k_BT_2}\right)-1\right]^{-1}= 1 +
2n_2(\omega^{\prime}), \label{d3}
\end{equation}
which will result in the friction force given by Eq.
\eqref{2three}. For propagating waves
$\mathrm{sgn}(\omega^{\prime})=\mathrm{sgn}(\omega)$ (for $\omega
>0$)  and from Eq. \eqref{2three} it follows that the contribution to
friction from propagating waves is equal to zero at zero
temperature, which agrees with the principle of relativity. However,
the contribution from evanescent waves is not equal to zero even
at zero temperature because in this case
$\mathrm{sgn}(\omega^{\prime})<0$ for $\omega < q_xV$. PL claim
that maybe the Lorentz transformation for the stress tensor is not
valid for the contribution to stress tensor from quantum
fluctuations. However, VP, instead of using Lorentz transformation
for the stress tensor, apply this transformation for calculation of
the electromagnetic field. This electromagnetic field is used to
calculate the stress tensor in the rest reference frame of plate
\textbf{1}. The result was the same. Thus contrary to the opinion
of PL, we argue that the Lorentz invariance exist only for quantum
fluctuations corresponding to propagating waves. This means that
the spectral characteristics of the electromagnetic field in
absolute vacuum (without of any bodies) are the same in all
inertial reference frames; otherwise it will contradict to the
principle of relativity. This result follows from the Lorentz
transformation for the electromagnetic field corresponding to the
propagating electromagnetic waves. For evanescent waves there is
no Lorentz invariance of the spectral properties of the
electromagnetic field. This result, which also follows from the
Lorentz transformation, does not contradict to the principle of
relativity because there are no evanescent waves in absolute
vacuum.

At zero temperature the integration in Eq. \eqref{2three} includes
only the interval $0< \omega <q_xV$. This integration takes into
account the contribution to friction from excitations in this
frequency range, which exist even at zero temperature. PL did not include
these excitations, and as a result they got zero friction.
Excitations which exist even at zero temperature contribute not
only to the lateral force, but also to normal force (see Eq.
\eqref{2five}). Thus the conservative Casimir-Lifshitz force also
contains some additional terms which were overlooked by PL.

Recently Pendry\cite{Pendry10} has also showed that the friction
is finite even at zero temperature, in qualitative agreement with
most previous approaches to the problem, but in contradiction to
the conclusion of PL. However Pendry considered a very simple
non-retarded and non-relativistic model. In contrast to Pendry, in
the framework of the same model we show that the calculation of PL
is in error. We show that this error is due to the assumption that
the zero-point radiation, corresponding to  evanescent
electromagnetic waves, obey Lorentz invariance. We also show that
normal component of Casimir-Lifshitz force calculated by PL is
also incorrect.

Since theory from Ref.\cite{Philbin1}is a limiting case of theory
from Ref.\cite{Philbin2}, our Comment is applicable for both
these papers.

\vskip 0.5cm

A.I.V acknowledges financial support from the Russian Foundation
for Basic Research (Grant N 10-02-00297-a) and ESF within activity
``New Trends and Applications of the Casimir Effect''.


\vskip 0.5cm







\begin{thebibliography}{999}

\bibitem{Rytov53}  S. M. Rytov, \textit{Theory of Electrical
Fluctuation and Thermal Radiation} (Academy of Science of USSR
Publishing, Moscow, 1953)

\bibitem{Rytov67}  M. L. Levin, S. M. Rytov, \textit{Theory of eqilibrium thermal
fluctuations in electrodynamics} (Science Publishing, Moscow,
1967)

\bibitem{Rytov89}  S. M. Rytov, Yu. A. Kravtsov, and V. I. Tatarskii, \textit{
Principles of Statistical Radiophyics} (Springer, New York.1989),
Vol.3




\bibitem{Lifshitz54}  E. M. Lifshitz, Zh. Eksp. Teor. Fiz. \textbf{29 }94
(1955) [Sov. Phys.-JETP \textbf{2 }73 (1956)]



\bibitem{Polder}  D. Polder and M. Van Hove, Phys. Rev. B \textbf{4}, 3303
(1971)



\bibitem{Volokitin99}  A. I. Volokitin and B. N. J. Persson, J.Phys.:
Condens. Matter \textbf{11}, 345
(1999);Phys.Low-Dim.Struct.\textbf{7/8},17 (1998)

\bibitem{VolokitinRMP07} A. I. Volokitin and B. N. J. Persson, Rev. Mod. Phys. \textbf{79}, 1291 (2007)

\bibitem{VolokitinUFN07} A.I.Volokitin and B.N.J.Persson, Usp. Fiz.
Nauk \textbf{177}, 921 (2007)[Phys. Usp. \textbf{50}, 879 (2007)]

\bibitem{Volokitin08}  A. I. Volokitin and  B. N. J. Persson, Phys. Rev. B \textbf{
78}, 155437 (2008)






\bibitem{Pitaevskii08} M. Antezza, L. P. Pitaevskii, S. Stringari,
and V. B. Svetovoy, Phys. Rev. A \textbf{77}, 022901 (2008)

\bibitem{Antezza05} M. Antezza, L. P. Pitaevskii, and S. Stringari,
Phys. Rev. Lett.  \textbf{95}, 113202 (2005)

\bibitem{Antezza06} M. Antezza, L. P. Pitaevskii, S. Stringari, and V. B. Svetovoy,
 Phys. Rev. Lett.  \textbf{97}, 223203 (2006)






\bibitem{Pendry99}  J. B. Pendry, J.Phys.:Condens.Matter \textbf{11}, 6621
(1999).

\bibitem{Volokitin01a}  A. I. Volokitin and B. N. J. Persson, Phys. Rev. B \textbf{
63}, 205404 (2001); Phys. Low-Dim. Struct. \textbf{5/6}, 151
(2001)

\bibitem{Volokitin04}  A. I. Volokitin and B. N. J. Persson, Phys. Rev. B \textbf{%
69}, 045417 (2004)

\bibitem{Volokitin03}  A. I. Volokitin and B. N. J.Persson, JETP Lett. \textbf{78}, 457 (2003)

\bibitem{Mulet01}  J. P. Mulet, K. Joulin, R. Carminati, and J. J. Greffet, Appl.
Phys.Lett. \textbf{78}, 2931 (2001)

\bibitem{Mulet02}  J. P. Mulet, K. Joulain, R. Carminati, and J. J. Greffet, Microscale
Thermophysical Engineering, \textbf{6}(3), 209 (2002)

\bibitem{Kittel}  A. Kittel, W. M\"uller-Hirsch, J. Parisi, S. Biehs,
D. Reddig, and M. Holthaus, Phys. Rev. Lett. \textbf{95}, 224301
(2005).

\bibitem{Joulain05}  K. Joulain, J. P. Mulet, F. Marquier, R. Carminati, and
J. J. Greffet, Surf. Sci. Rep. , \textbf{57, }59 (2005)











\bibitem{Volokitin01b}  A. I. Volokitin and B. N. J. Persson, J.Phys.:
Condens. Matter \textbf{13}, 859 (2001)

\bibitem{Volokitin03a}  A. I. Volokitin and B. N. J. Persson, Phys. Rev. Lett.
\textbf{91}, 106101 (2003)

\bibitem{Volokitin03b}  A. I. Volokitin and B. N. J. Persson, Phys. Rev. B,
\textbf{68}, 155420 (2003)

\bibitem{Volokitin06}  A. I. Volokitin and  B. N. J. Persson, Phys. Rev. B \textbf{
74}, 205413 (2006)










\bibitem{Philbin1} T. G. Philbin and U. Leonhardt, New J. Phys.
\textbf{11}, 033035 (2009)

\bibitem{Philbin2} T. G. Philbin and U. LeonhardtG, arXiv:0904.2148v3 [quant-ph],
2009












\bibitem{Pendry97}  J.B. Pendry, J. Phys. C\textbf{9} 10301 (1997)


\bibitem{Volokitin10}  A. I. Volokitin and  B. N. J. Persson, Phys. Rev.
 \textbf{81}, 239901(E) (2010)

\bibitem{Lifshitz1961} I.E. Dzyaloshinskii, E.M. Lifshitz and L.P.
Pitaevskii, Adv. Phys. \textbf{10}, 165 (1961)

\bibitem{Lifshitz1980} L.D. Landau, E.M. Lifshitz and L.P.
Pitaevskii, textit{Statistical Physics, Part 2}
(Butterworth-Heinemann, Oxford, 1980)

\bibitem{Pendry10}  J.B. Pendry,  New J. Phys.
\textbf{12}, 033028 (2010)










\end{thebibliography}
\end{document}